# MULTIDIMENSI PADA DATA WAREHOUSE DENGAN MENGGUNAKAN RUMUS KOMBINASI

**Spits Warnars Harco Leslie Hendric**
*Fakultas Teknologi Informasi, Universitas Budi Luhur*
*E-mail: spits@bl.ac.id*

**ABSTRACT**

*Multidimensional in data warehouse is a compulsion and become the most important for information delivery, without multidimensional data warehouse is incomplete. Multidimensional give the able to analyze business measurement in many different ways. Multidimensional is also synonymous with online analytical processing (OLAP).*

## 1. PENDAHULUAN

Tabel Fakta terbentuk berdasarkan hypercubes merupakan kumpulan lebih dari 3 dimensi yang terbentuk dari analisa dari sebuah laporan. Jika hanya 2 atau 3 dimensi maka tabel fakta terbentuk berdasarkan cube. Dengan perintah *structured query language* (sql) sederhana yang mengakses tabel fakta dihasilkan sebuah tabel database temporary yang akan bernilai maknanya jika ditampilkan dalam bentuk grafik. Bentuk grafik ini adalah gambaran tampilan data yang secara kasat mata akan lebih mudah untuk dipahami dalam menentukan suatu keputusan manajemen dibandingkan dengan tampilan data yang bersifat angka. Beberapa peneliti mengatakan bahwa gambar mempunyai banyak makna dibanding angka, selain itu akan lebih mudah melihat suatu tren penurunan atau kenaikan melalui gambar grafik dibandingkan dengan data angka.

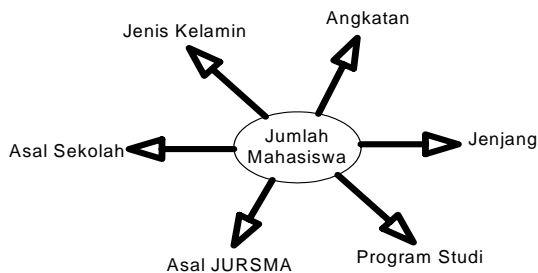

**Gambar 1.** Hypercubes

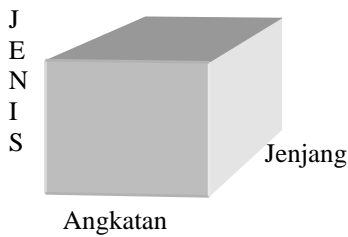

**Gambar 2.** Cubes

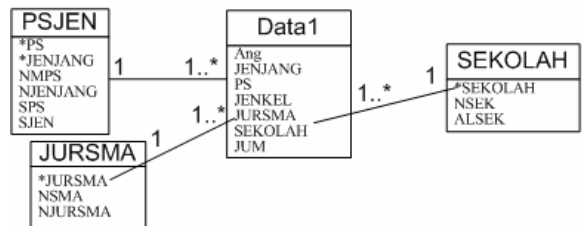

**Gambar 3.** Tabel Fakta Data1 dan table dimensi PSJEN, JURSMA dan SEKOLAH

Sebagai contoh perhatikan tabel database fakta dibawah ini:

**Tabel 1.** Contoh database fakta

| Angkatan | Jen | PS | JENIS | jum |
|---|---|---|---|---|
| 2000 | 5 | 11 | p | 11 |
| 2000 | 5 | 11 | w | 22 |
| 2000 | 3 | 11 | p | 12 |
| 2000 | 3 | 11 | w | 13 |
| 2000 | 5 | 22 | p | 10 |
| 2001 | 5 | 11 | w | 33 |
| 2001 | 5 | 11 | p | 44 |
| 2001 | 3 | 11 | w | 14 |
| 2001 | 3 | 11 | p | 15 |
| 2002 | 5 | 11 | p | 55 |
| 2002 | 5 | 11 | w | 66 |
| 2002 | 3 | 11 | p | 16 |
| 2002 | 3 | 11 | w | 17 |

dengan memberikan perintah sql seperti dibawah ini:

```
    select ang,jenj,jenkel, sum(jum) as
jumlah
    from dwmhs where ang="2000"
        group by ang,jenj,jenkel;
```

Maka hasil dari perintah sql diatas akan menampilkan laporan seperti dibawah ini:

```
Angkatan: 2000
```
| Jenis | S1 | D3 |
|---|---|---|
| Pria | 21 | 12 |
| Wanita | 22 | 13 |

Jika laporan tersebut ditampilkan dalam bentuk grafik akan berbentuk seperti dibawah ini:

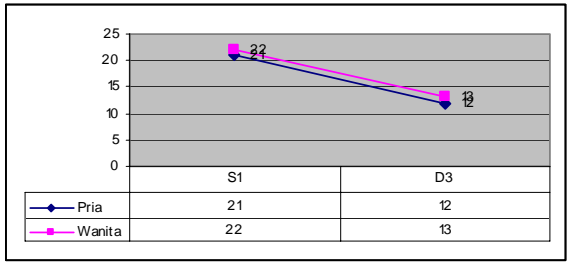

**Gambar 3.** Bentuk grafik dari laporan hasil perintah sql

## 2. SLICE DAN DICE

Kemampuan multidimensi hypercubes atau cube dalam memberikan informasi dalam pengambilan keputusan dapat ditingkatkan dengan menggunakan konsep slice dan dice. Konsep slice dan dice pada data warehouse ini merupakan sebuah konsep multi dimensi pada datawarehouse, dimana hypercubes atau cube dapat dilihat dari berbagai dimensi. Selain itu konsep slice dan dice berfungsi untuk mengambil potongan hypercubes atau cube berdasarkan nilai tertentu pada satu atau beberapa dimensinya. Konsep slice dan dice ini dapat dilakukan dengan memberikan query atau perintah *structured query language* (sql) sederhana.

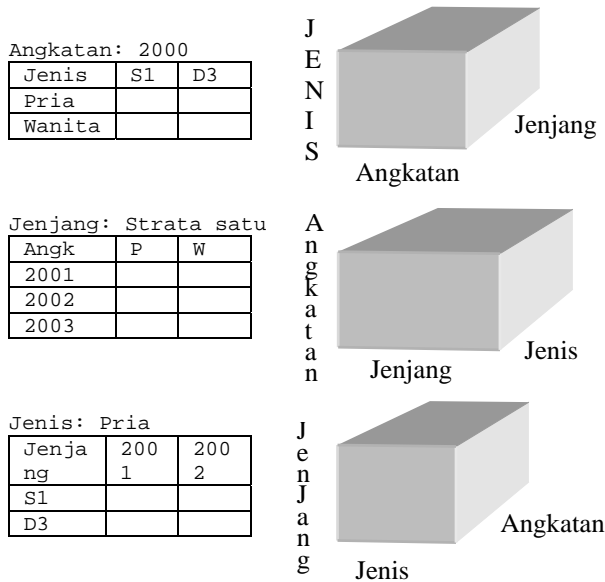

**Gambar 4.** Konsep Slice dan Dice

## 3. ROLL UP DAN DRILL DOWN

Selain menggunakan slice dan dice kemampuan multidimensi hypercubes atau cube dalam memberikan informasi dalam pengambilan keputusan dapat ditingkatkan dengan menggunakan konsep roll up dan drill down. Roll up adalah proses generalisasi satu atau beberapa dimensi dengan merangkum atau meringkas nilai-nilai ukurannya. Dengan kata lain generalisasi berarti naik ke tingkat atasnya dalam hirarki dimensi. Sedangkan proses drill down adalah proses memilih dan menampilkan data rincian dalam satu atau beberapa dimensi dan merupakan kebalikan dari operasi roll-up.

Sama seperti konsep slice dan dice dapat dilakukan dengan memberikan query atau perintah *structured query language* (sql) sederhana, demikian juga dengan konsep roll up dan drill down dapat dilakukan dengan memberikan query atau perintah *structured query language* (sql) sederhana.

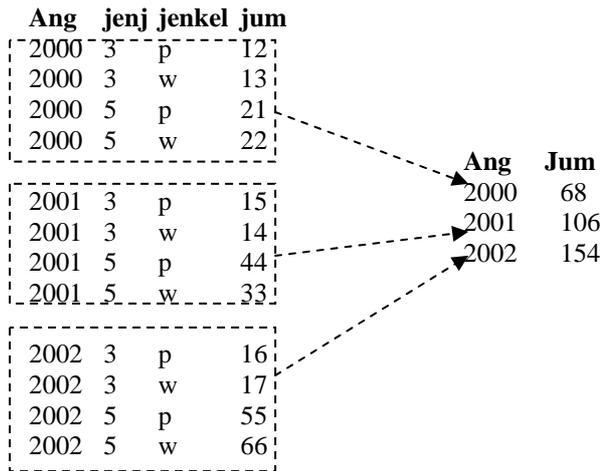

**Gambar 5.** Konsep Roll up

Konsep roll up pada gambar 5 diatas dalam menggeneralisasi data dilakukan dengan perintah sql dibawah ini:

```
Select Ang, sum(jum) as jum from DWmhs group by ang
```

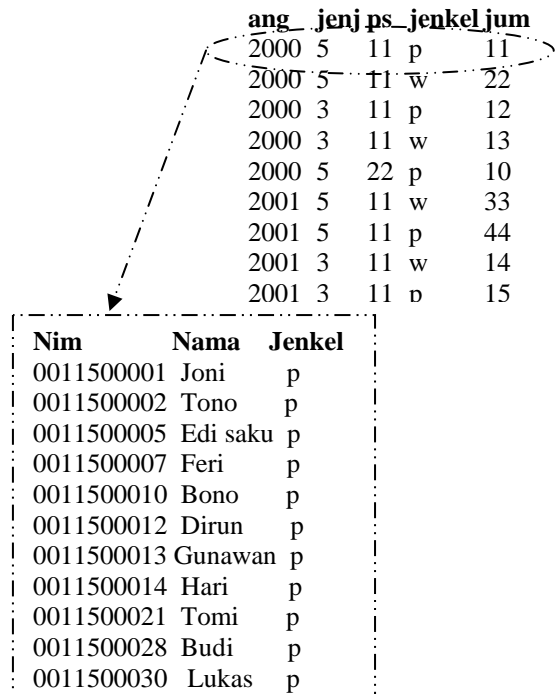

**Gambar 6.** Konsep Drill down

Konsep drill down pada gambar 6 diatas dalam merinci data dilakukan dengan perintah sql dibawah ini:

```
Select a.nim,a.nama from mastmhs a, DWmhs b
   where left(a.nim,2)=right(b.ang,2) and
            substr(a.nim,3,2)=b.ps and
      substr(a.nim,5,1)b.jenj and
a.jenkel=b.jenkel
```

Manajemen pengambil keputusan akan sangat terbantu dalam mengambil keputusan dengan melihat hypercubes atau cube secara generalisasi dengan konsep roll up dan secara terinci dengan konsep drill down. Generalisasi data dengan konsep roll up membantu manajemen pengambil keputusan dengan data-data yang bersifat rangkuman atau ringkasan. Rincian data dengan konsep drill down membantu manajemen pengambil keputusan dengan data-data yang bersifat terinci.

## 4. PERTANYAAN-PERTANYAAN

Berdasarkan uraian-uraian diatas terlihat bahwa konsep multidimensi dengan menerapkan konsep slice dan dice, roll up dan drill down akan meningkatkan kemampuan hypercubes atau cube dalam memberikan informasi pengambilan keputusan bagi manajemen pengambil keputusan. Meningkatnya kemampuan hypercubes atau cube dalam memberikan informasi pengambilan keputusan, menimbulkan pertanyaan-pertanyaan yang akan dibahas lebih lanjut pada pembahasan berikutnya. Adapun pertanyaan-pertanyaan tersebut adalah:

1. Berapa dimensi minimal dan dimensi maksimal yang dapat dibentuk dari sebuah hypercubes atau cube?
2. Berapa kombinasi laporan atau grafik yang dapat dibentuk dari sebuah hypercubes atau cube?
3. Berapa kombinasi laporan atau grafik yang dapat dibentuk pada setiap dimensi dari sebuah hypercubes atau cube?

## 5. PEMBAHASAN AWAL

Laporan yang ditampilkan dalam bentuk sebuah grafik merupakan perpotongan antara sumbu horisontal dan sumbu vertikal. Sumbu horisontal pada sebuah grafik hanya menggambarkan nilai sebuah kolom, sedangkan sumbu vertikal menggambarkan nilai sebuah kolom atau perpaduan lebih dari satu kolom. Kolom bisa juga diartikan sebagai dimensi pada hypercubes atau cube.

Maksimal kolom yang dapat dipakai sebagai sumbu horisontal adalah jumlah kolom tabel fakta (atau jumlah dimensi pada hypercubes atau cube) selain kolom jumlah. Kolom jumlah tidak dapat dijadikan sebagai sumbu vertikal maupun sumbu horisontal dikarenakan kolom jumlah adalah nilai yang akan ditampilkan pada perpotongan sumbu vertikal dan sumbu horisontal pada tampilan grafik.

Kolom jumlah ini akan memperlihatkan nilai tren kenaikan atau penurunan pada sebuah tampilan grafik.

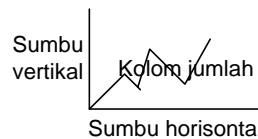

**Gambar 7.** Grafik

Sumbu vertikal adalah dimensi yang dapat dibentuk dari sebuah kolom/dimensi atau perpaduan nilai kolom/dimensi. Sehingga kita dapati maksimal nilai dimensi suatu tabel fakta adalah jumlah kolom tabel fakta tersebut selain kolom jumlah. Sehingga apabila sebuah tabel mempunyai 2 kolom selain kolom jumlah maka akan mempunyai maksimal 2 dimensi, jika mempunyai 3 kolom selain kolom jumlah maka akan mempunyai maksimal 3 dimensi dan seterusnya.

Hal ini juga dapat terlihat pada hypercubes atau cube dimana jumlah dimensi pada hypercubes atau cube adalah sama dengan jumlah kolom pada tabel fakta selain kolom jumlah. Perhatikan gambar dibawah ini dimana jumlah dimensi pada hypercubes yaitu 6 sama dengan jumlah field pada tabel fakta data1 selain field jum yaitu 6.

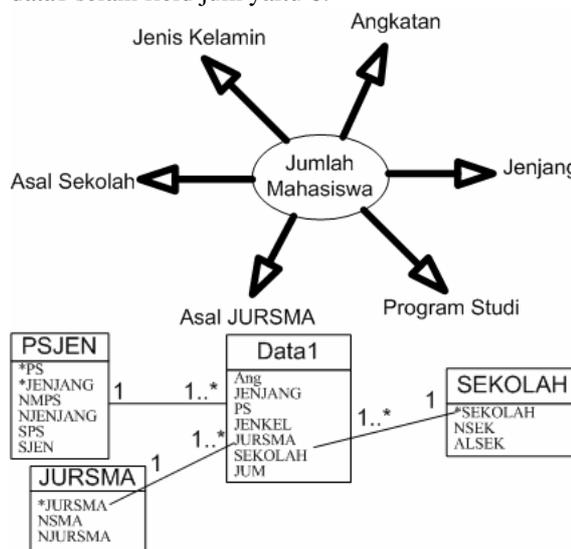

**Gambar 8.** Hypercubes dan tabel fakta Data1

Dengan demikian pertanyaan berapa dimensi minimal dan dimensi maksimal yang dapat dibentuk dari sebuah hypercubes atau cube telah terjawab. Dimana jumlah dimensi minimal adalah 1 dan jumlah dimensi maksimal adalah sebanyak dimensi yang dimiliki oleh hypercubes atau cube.

## 6. RUMUS KOMBINASI

Untuk mendapatkan kombinasi laporan/tabel yang dapat dibentuk dari masing-masing dimensi pada hypercubes tanpa memperhatikan urutan susunannya serta untuk menjawab pertanyaan berapa kombinasi laporan atau grafik yang dapat dibentuk dari sebuah tabel fakta maka ada baiknya kita menggunakan rumus kombinasi pada ilmu statistik.

Teori kombinasi mengatakan "*Kombinasi dari sejumlah n objek yang berbeda yang diambil sejumlah r pada satu saat adalah pemilihan r objek itu tanpa memperhatikan urutan susunannya*". (Jong Jek Siang, 2002).

Jumlah kombinasi dari objek sejumlah n yang diambil r pada satu saat mempunyai rumus sebagai berikut:

nCr atau C(n,r) atau C n,r atau $\binom{n}{r}$

dimana: $n\,C\,r = \dfrac{n!}{r!\,(n-r)!}$

dimana: $n! = n(n-1)(n-2)\ldots 1$

maka:
$0! = 1$
$1! = 1$
$2! = 2*1 = 2$
$3! = 3*2*1 = 6$
$6! = 6*5*4*3*2*1 = 720$

Dengan menggunakan rumus kombinasi diatas maka dikembangkan sebuah rumus yang dapat memperlihatkan jumlah laporan atau grafik yang dapat dibentuk berdasar jumlah dimensi pada hypercubes untuk memenuhi konsep multi dimensi pada datawarehouse yaitu:

$n * n{-}1\,C\,r{-}1$

dimana:

n adalah jumlah dimensi hypercubes
r adalah nilai dimensi yang akan dibentuk

Dengan dikembangkannya rumus ini maka kita akan mengetahui:
1. Keseluruhan kombinasi laporan atau grafik yang dapat dibentuk
2. Kombinasi laporan atau grafik pada setiap dimensi

Timbul pertanyaan lanjutan: kenapa pembahasan dalam rangka mendapatkan dimensi pada sebuah hypercubes atau cube dikembangkan dari rumus teori kombinasi statistik dan kenapa tidak menggunakan rumus teori lainnya ? Untuk menjawab pertanyaan tersebut ada baiknya kita bahas beberapa alasan yang mendasari pemilihan rumus teori kombinasi tersebut.

Untuk pembuatan laporan dimensi 2 keatas kolom pertama bersifat tetap sedangkan kolom berikutnya selain kolom jumlah dapat ditukar posisinya. Kolom pertama tidak dapat dipertukarkan dikarenakan kolom pertama ini menjadi nilai pada garis horisontal pada grafik, sedangkan kolom jumlah adalah nilai yang akan ditampilkan pada perpotongan baris dan kolom pada tampilan grafik.

Sebagai contoh perhatikan penjelasan perbandingan tampilan laporan dapat dilihat di tabel 2.

**Tabel 2.** Contoh penjelasan perbandingan tampilan laporan

| Angkatan | Jenjang | Jenis | Jumlah |
|---|---|---|---|
| 2000 | 5 | P | 21 |
| 2000 | 5 | W | 22 |
| 2000 | 3 | P | 12 |
| 2000 | 3 | W | 13 |
| 2001 | 5 | P | 44 |
| 2001 | 5 | W | 33 |
| 2001 | 3 | P | 15 |
| 2001 | 3 | W | 14 |
| 2002 | 5 | P | 55 |
| 2002 | 5 | W | 66 |
| 2002 | 3 | P | 16 |
| 2002 | 3 | W | 17 |
| Jumlah | | | **328** |

Laporan di atas jika ditampilkan dalam bentuk grafik akan mempunyai tampilan grafik seperti dibawah ini:

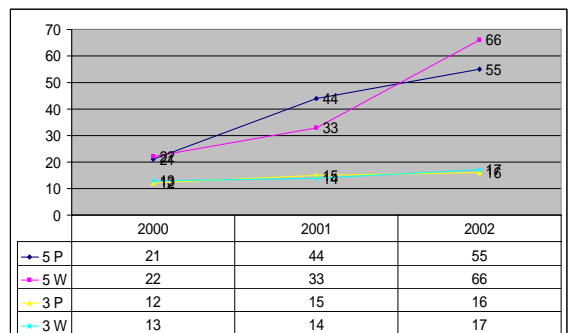

**Gambar 9.** Tampilan grafik hasil laporan

Tampilan laporan/tabel diatas dapat mempunyai tampilan laporan yang ditukar posisi kolomnya selain kolom pertama dan kolom jumlah, yaitu kolom Jenis dipindah ke posisi kolom Jenjang sebaliknya kolom Jenjang dipindah ke posisi kolom Jenis. Sehingga tampilan laporan diatas akan berubah menjadi laporan seperti dibawah ini:

**Tabel 3.** Contoh penjelasan perbandingan tampilan laporan

| Angkatan | Jenis | Jenjang | Jumlah |
|---|---|---|---|
| 2000 | P | 5 | 21 |
| 2000 | W | 5 | 22 |
| 2000 | P | 3 | 12 |
| 2000 | W | 3 | 13 |
| 2001 | P | 5 | 44 |
| 2001 | W | 5 | 33 |
| 2001 | P | 3 | 15 |
| 2001 | W | 3 | 14 |
| 2002 | P | 5 | 55 |
| 2002 | W | 5 | 66 |
| 2002 | P | 3 | 16 |
| 2002 | W | 3 | 17 |
| Jumlah | | | **328** |

Laporan diatas jika ditampilkan dalam bentuk grafik akan mempunyai tampilan grafik seperti dibawah ini:

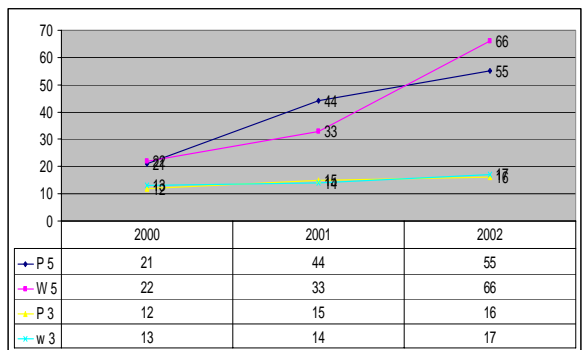

**Gambar 10.** Bentuk grafik hasil laporan tabel 3

Terlihat bahwa tampilan kedua laporan/tabel diatas tidak berbeda dalam hal nilai pada kolom jumlah, yang berbeda hanyalah adanya penukaran kolom jenis dan jenjang. Demikian juga dengan tampilan kedua grafik diatas terlihat bahwa bentuk grafik sama, mempunyai garis horizontal yang berisi angkatan yang tetap sama. Yang berbeda hanyalah tampilan pada legenda grafik yaitu untuk legenda garis warna biru pada grafik pertama adalah 5 p, sedang pada grafik kedua adalah p 5.

Dari penjelasan perbandingan bentuk kedua laporan tersebut jelas bahwa kombinasi dimensi yang akan dicari adalah kombinasi sejumlah n objek yang berbeda yang diambil dari sejumlah r tanpa memperhatikan urutan susunannya. Hal ini sejalan dengan isi dan penjelasan rumus teori kombinasi, sehingga pembahasan untuk mendapatkan dimensi pada sebuah hypercubes atau cube digunakan rumus teori kombinasi statistik.

## 7. KOMBINASI RUMUS 3 DIMENSI

Untuk mencoba pembuktian pengembangan rumus kombinasi diatas maka dipandang perlu untuk menguji coba pengembangan rumus kombinasi tersebut dengan mengambil sampel dari sebuah hypercubes atau cube 3 dimensi. Untuk memudahkan penjelasan maka setiap dimensi akan diwakilkan dengan huruf alphabet sehingga hypercubes atau cube 3 dimensi tersebut mempunyai tampilan laporan sebagai berikut:

| A | B | C | Jumlah |
|---|---|---|---|
|   |   |   |   |
|   |   |   |   |

Dengan menggunakan rumus kombinasi:
   $n * n-1 \, C \, r-1$

dimana n adalah jumlah dimensi hypercubes dan r adalah nilai dimensi yang akan dibentuk. Karena jumlah dimensi hypercubes contoh adalah 3 maka n=3. Sesuai dengan konsep multidimensi data warehouse maka berdasarkan hypercubes 3 dimensi ini akan dicari kombinasi laporan/tabel atau grafik yang dapat dibentuk pada setiap dimensinya.

Dimensi 1, berarti r=1          Dimensi 2, berarti r=2  
maka $= n * n-1 \, C \, r-1$     maka $= n * n-1 \, C \, r-1$  
$\quad = 3 * 3-1 \, C \, 1-1$     $\quad = 3 * 3-1 \, C \, 2-1$  
$= 3 * 2 \quad C \, 0$            $= 3 * 2 \quad C \, 1$  
$= 3 * (\frac{n!}{r!(n-r)!})$     $= 3 * (\frac{n!}{r!(n-r)!})$  
$= 3 * (\frac{2!}{0!(2-0)!})$     $= 3 * (\frac{2!}{1!(2-1)!})$  
$= 3 * (\frac{2!}{0!(2!)})$       $= 3 * (\frac{2!}{1!(1!)})$  
$= 3 * (\frac{1*2}{1(1*2)})$      $= 3 * (\frac{1*2}{1(1)})$  
$\quad = 3 * 1$                   $\quad = 3 * 2$  
$= 3$                             $= 6$

Dimensi 3, berarti r=3  
maka $= n * n-1 \, C \, r-1$  
$\quad = 3 * 3-1 \, C \, 3-1$  
$= 3 * 2 \quad C \, 2$  
$= 3 * (\frac{n!}{r!(n-r)!})$  
$= 3 * (\frac{2!}{2!(2-2)!})$  
$= 3 * (\frac{2!}{2!(0!)})$  
$= 3 * (\frac{1*2}{1*2(1)})$  
$= 3 * 1$  
$= 3$

Hasil dari rumus kombinasi ini pada hypercubes 3 dimensi ini dapat terlihat pada tabel dibawah ini:

**Tabel 4.** Hasil rumus kombinasi hypercubes 3 dimensi

| Sumbu horizontal | Dimensi |    |             |
|---|---|---|---|
|   | 1 | 2 | 3 |
| A | A | AB AC | A BC atau A CB |
| B | B | BA BC | B AC atau A CA |
| C | C | CA CB | C AB atau A BA |
| Kombinasi | 3 | 6 | 3 |
| Total: 12 kombinasi ||||

Pada tabel diatas pada dimensi 3 kombinasi laporan/tabel/grafik mempunyai 2 pilihan misalnya untuk sumbu horisontal A dapat mempunyai tampilan laporan sebagai berikut:

| A | B | C | Jumlah |
|---|---|---|---|
|   |   |   |   |
|   |   |   |   |

Sesuai dengan penjelasan sebelumnya bahwa untuk pembuatan laporan dimensi 2 keatas kolom pertama bersifat tetap sedangkan kolom berikutnya selain kolom jumlah dapat ditukar posisinya, , dan kolom pertama ini menjadi sumbu horizontal pada tampilan grafik. Sehingga laporan diatas dapat dirubah menjadi seperti laporan dibawah ini dan mempunyai makna yang sama.

| A | C | B | Jumlah |
|---|---|---|---|
|   |   |   |   |
|   |   |   |   |

Terlihat dari pembuktian dengan rumus kombinasi n* n-1 C r-1 diatas:
pada saat dimensi 1 yaitu r=1 menghasilkan 3 kombinasi laporan/tabel/grafik
pada saat dimensi 2 yaitu r=2 menghasilkan 6 kombinasi laporan/tabel/grafik
pada saat dimensi 3 yaitu r=3 menghasilkan 3 kombinasi laporan/tabel/grafik
Sehingga total tampilan laporan/tabel/grafik yang dapat dibentuk adalah 12 kombinasi

Perpotongan antar 2 dimensi dibawah ini akan memperlihatkan adanya 6 perpotongan kolom yaitu AB, AC, BA, BC, CA dan CB.

|   | A | B | C |
|---|---|---|---|
| A |   | AB | AC |
| B | BA |   | BC |
| C | CA | CB |   |

Hal ini sesuai dengan hasil pembuktian dengan rumus pada saat dimensi 2 akan menghasilkan 6 kombinasi laporan/tabel/grafik.

Perpotongan antar 3 dimensi dibawah ini akan memperlihatkan adanya 6 perpotongan kolom yaitu A BC, A CB, B AC, B CA, C AB dan C BA.

|   | A | | B | | C | |
|---|---|---|---|---|---|---|
|   | B | C | A | C | A | B |
| A |   |   |   | A BC |   | A CB |
| B |   | B AC |   |   | B CA |   |
| C | C AB |   | C BA |   |   |   |

Sesuai dengan penjelasan sebelumnya bahwa untuk pembuatan laporan dimensi 2 keatas kolom pertama bersifat tetap sedangkan kolom berikutnya selain kolom jumlah dapat ditukar posisinya, , dan kolom pertama ini menjadi sumbu horizontal pada tampilan grafik. Oleh karena itu 6 perpotongan kolom diatas karena mempunyai bentuk tampilan laporan yang sama dan mempunyai bentuk tampilan grafik yang sama dapat dipertukarkan kolom berikutnya selain kolom pertama dan kolom jumlah.

Jadi oleh karena 6 perpotongan kolom diatas mempunyai kolom pertama yang sama dan kolom berikutnya yang sama yang saling dipertukarkan posisinya yaitu:

      A BC   atau A CB
      B AC   atau B CA
      C AB   atau C BA

Sehingga sebenarnya hanya ada 3 perpotongan kolom dan hal ini sesuai dengan hasil pembuktian dengan rumus pada saat dimensi 3 akan menghasilkan 3 kombinasi laporan/tabel/grafik.

Terlihat juga dengan pembuktian rumus kombinasi pada hypercubes 3 dimensi ini kombinasi awal dan akhir mempunyai nilai yang sama dengan jumlah dimensi yaitu 3.

## 8. KESIMPULAN

Akhirnya kita akan mengambil kesimpulan bahwa:
1. Nilai kombinasi awal dan akhir mempunyai nilai yang sama dengan jumlah dimensi selain kolom jumlah pada tabel fakta atau hypercubes
2. Rumus kombinasi ini dapat mempermudah dan menjadi acuan dalam membuat sebuah aplikasi OLAP (Online Analytical Processing) yang mengakses data warehouse dan dapat menampilkan kemampuan mutidimensi dari sebuah hypercubes atau cube secara lebih maksimal.
3. secara konsep perintah sql yang digunakan untuk mengakses hypercubes data warehouse mempunyai kesamaan sebagai berikut:

```
select    field1..fieldn,   sum(jum)   as
jumlah
    from namaTabel
        group by field1..fieldn;
```

Dimana urutan `select field1…fieldn` harus sama dengan `group by field1…fieldn`

## DAFTAR PUSTAKA